\documentstyle[version2,preprint,aps]{revtex}  % for REVTeX 3.0
  \newcounter{mycount}
  \def\aea{\setcounter{mycount}{\value{equation}}\stepcounter{mycount}
           \setcounter{equation}{0}
           \def\theequation{\arabic{mycount}\alph{equation}}
           \begin{eqnarray}}
  \def\zea{\end{eqnarray}\setcounter{equation}{\value{mycount}}
           \def\theequation{\arabic{equation}}}
  \let\qd=\quad   \let\de=\delta
  \let\e=\varepsilon  \let\h=\eta \let\k=\kappa \let\la=\lambda
  \let\r=\rho \let\s=\sigma \let\G=\Gamma \let\D=\Delta
  \def\0{\over } \def\2{{1\over2}} \def\6{\partial } \def\9{\dot }
  \def\<{\langle } \def\>{\rangle } \def\lb{\left} \def\rb{\right}
  \let\ap=\approx \let\eq=\equiv \let\bl=\biggl \let\br=\biggr
  \def\bea{\begin{eqnarray}} \def\eea{\end{eqnarray}}
  \def\beq{\begin{equation}} \def\eeq{\end{equation}}
\begin{document}
\draft
\begin{title}
 $^3$He $A\to B$ Transition in Finite Geometries
\end{title}
\author{Peter Kost\"adt and Mario Liu}
\begin{instit}
 Institut f\"ur Theoretische Physik, Universit\"at Hannover,
 30167 Hannover, Germany
\end{instit}
\bigskip\bigskip
\noindent\hspace*{7mm} PACS. 67.50F\ -\ Superfluid phase \\
         \hspace*{7mm} PACS. 68.10J\ -\ Kinetics
\begin{abstract}
 {\bf Abstract}.\ -\
 The measured and reported rate \cite{boyd92} of the $^3$He $A\!\to\! B$
 transition depends on the diameter of the experimental cell. Theoretically,
 this can only be explained by a direct drag on the interface exerted by the
 lateral walls, leading to a pressure discontinuity across the interface.
 This has not been considered before, either here or in similar circumstances.
 The general hydrodynamic theory that includes the influence of finite
 geometries is presented.
\end{abstract}
\newpage
%\tightenlines
{\bf Introduction}.\ -\
General hydrodynamic considerations~\cite{graliu90,paliu92,koliu93} on the
$^3$He $A\to B$ transition have in the past yielded a fairly complete
picture for the infinite geometry. Depending on the interface velocity
$\9u$, there are two qualitatively different types of behavior.
If $\9u$ is small compared to the second-sound velocity,
$\9u\ll c_2$, three characteristic features of the {\it phase coherent
transition} are predicted: (i) The chemical potential is continuous across
the interface, forcing the melting $A$ phase to be warmer than the growing
$B$ phase. (ii) The relevant mode of heat transfer, second sound, is
efficient enough to obliterate any onset of hypercooling. (iii) The effective
growth resistance preventing the transition rate to diverge consists of three
elements in series: two account for the sq-mode, a hydrodynamically gentle
temperature variation on each side of the interface, the third
accounts for a sharp temperature discontinuity at the microscopically narrow
interface. The former two dissipative elements stem from collisions among
quasiparticles, the latter one (probably much smaller in magnitude) results
from Andreev scattering of these particles at the interface.
At the fast end of the transition, $\9u\gg c_2$, circumstances are rather
different: (i) Second sound is now a comparatively slow process, leaving the
chemical potential discontinuous and the growing $B$ phase warmer, since an
ever-increasing portion of the latent heat is no longer removed from the
interface region. (ii) The usual scenario of hypercooling (in which no heat
at all is transfered) becomes asymptotically valid. (iii) The total growth
resistance contains both additive and multiplicative contributions.

Since second sound is so important in determining the properties of the
$A$-$B$ transition, it seems plausible that the above features would not easily
survive in a confining geometry: walls hold back the normal velocity
$v_n$ and thus confine the range of heat transfer via second sound.
In what follows we briefly state what this range is, why its finiteness is
important, and that it alone is insufficient to account for the
measured~\cite{boyd92} geometry dependence of the phase transition.
Generally, for stationary flows, lateral walls are well-accounted for
by a relaxation term $\r v_n/\tau$ in the Navier-Stokes equation
(obtained by averaging over the cross section of the experimental cell, eg.
$\tau=\r R^2/8\h$ for a cylinder of radius $R$). As a result of this term,
the second-sound step functions sent out by the moving interface
\cite{graliu90,paliu92,koliu93} acquire an extra, exponentially decaying
factor. In the rest frame of the interface, with $B$ phase at $x<0$ and $A$
phase at $x>0$, it is $\exp(x/\la_2)$. The resulting stationary temperature
profile is shown in Fig.~1 and 2. (On the scale of these figures, the
temperature variations of the sq- and diffusive modes~\cite{koliu93} are too
small to be separated from the discontinuity at the interface, $x=0$.)
The decay length
\beq \lambda_2=-\rho_n\tau(c_2^2-\dot u^2)/(\rho_s\dot u) \label{la2}\eeq
is also the range of heat transfer via second sound. Outside this range,
$|x|\gg |\la_2|$, there is no means for heat transfer, and energy balances
locally. Now, it is important to realize
that the usual thermodynamic consideration of hypercooling is based on
local balance of energy. Hence, the temperature $T^B=T^A+\de T$ of the growing
$B$ phase is taken to be given by the $A$ phase temperature $T^A$ and the
heating $\de T$ from the released latent heat. Second sound transfers heat
away from the interface and invalidates this consideration.
However, if the heat transfer has a finite range, hypercooling remains a valid
concept outside this range. And indeed, Fig.~1 shows that the $B$ phase is
warmer than the $A$ phase for $x\gg |\la_2|$. In addition, the solution
depicted in Fig.~1 is obtained under the condition
\beq \Bigl[|\D\mu|-(\D\s)^2/\6_T\s^B\Bigr]_{T=T_i}\geq 0\;, \label{HC}\eeq
which is, up to first order in $T_{AB}-T_i$, equivalent to the usual and
approximative hypercooling condition $(T_{AB}-T_i)-l_{AB}/c\geq 0$.
($T_i:$ initial undercooling temperature; $T_{AB}:$ coexistence
temperature; $\mu :$ chemical potential, $\s :$ entropy, $l_{AB}:$ latent
heat, $c:$ specific heat, all per unit mass; $\D:$ difference across the
interface, eg. $\D\mu\eq\mu^B-\mu^A\,$.)

Above the hypercooled region, $\e\eq 1-T_i/T_{AB}\alt 0.5\,\%$, planar
instability and non-stationarity (the usual conundrum of snow flakes) may be
expected to be relevant, but are not: Approaching the critical value
$\e_c\sim 0.5\,\%$ from below, $\9u$ vanishes with the left side of
Eq.~(\ref{HC}), and $\la_2$ diverges accordingly.
Once $\la_2$ exceeds the longitudinal dimensions of the experimental cell,
even a heat transfer within the scale of $\la_2$ is sufficient
to  disturb the energy balance and invalidate the concept of hypercooling.
Then a planar transition becomes viable again.
(The appropriate solution is non-stationary, similar to that of
Ref.~\cite{graliu90}; see discussion at the end of the introduction.)

Concentrating now on the hypercooled region, where the stationary solutions
of Fig.~1 and 2 are valid, we nevertheless face the following difficulty:
Despite the relaxation term $\r v_n/\tau$ and the resulting qualitatively
changed behavior of second sound, the interface velocity $\9u$ remains
independent of the cell radius $R$, in stark contrast with the
well-documented experimental observations~\cite{boyd92}. This is indeed
perplexing, since the hydrodynamic description accounts for the structure
of the dynamics and is independent of specific mechanisms, especially
if any has been overlooked. Re-examining the equations of
Ref.~\cite{graliu90,paliu92,koliu93}, we found no reason to doubt the
strict conservation of energy and mass, leaving the momentum
as the only candidate to resolve the puzzle: Rough lateral walls break the
translational symmetry and it is conceivable that the Navier-Stokes equation
acquires a relaxation term that is singular at the interface, resulting in a
discontinuity of the momentum current,
\begin{equation} \Delta p=\Gamma\rho\dot u/R\ . \label{Dp}\end{equation}
This equation can be viewed as the equation of motion for the interface, in
which the inertial term $\sim\ddot u$ and the restoring force $\sim u$
vanish; the pressure discontinuity $\Delta p$ is the external force
(per unit area), while the damping $\sim\dot u$ comes from the drag exerted
by the lateral walls and would have its origin, for instance, in the
interaction between the surface tension (including textural
contributions) and the wall's roughness. The total drag should be proportional
to the contact length, $2\pi R$ in a cylinder, per unit area it is $\sim 1/R$.
$\Gamma$ is introduced as a phenomenological parameter, accounting for the
degree of roughness. As we shall see, Eq.~(\ref{Dp}) indeed
establishes agreement between theory and experiment.
More specifically, we find a ``wall contribution'' $\sim 1/R$ added
to the total growth resistance of the infinite geometry, so that the change of
$\dot u$ is also proportional to $1/R$.
(The necessity of having $\D p\not=0$ has not arisen before, such as at the
well-studied $^4$He~II-vapor or $^4$He~II-solid interfaces~\cite{graliu88}.
At least, it was not studied there. A re-examination may be worthwhile.)

Finally, we address the relationship between the stationary solution of
second sound presented here and the propagating step functions of
Ref.~\cite{graliu90,paliu92,koliu93} in the case of an infinite geometry.
It is noteworthy that letting $R\!\to\!\infty$, the exponentially decaying
solution does not reduce to the same steps. This is because of the following
reasons: In deriving the stationary solution here, the lateral dimension is
taken to be finite, the longitudinal one infinite; and although an arbitrarily
large $R$ results in an arbitrarily large decay length $\la_2$, it is
finite and always smaller than the longitudinal dimension.
The energy current $Q$ in the present solution is constant, it always is
in stationary, one-dimensional solutions, $\6_xQ=-\6_te\eq 0$. Since $Q$
vanishes outside $\la_2$ (aside from a constant equilibrium contribution)
it must also vanish within \cite{seven}. This alone is enough
to exclude the type of solution represented by the step functions propagating
toward infinity, all the while transferring energy. The two types of solutions
are in fact mutually exclusive: If the lateral geometry is confining enough
to entail a decay length $\la_2$ much smaller than the longitudinal dimension,
the stationary solution presented here is relevant, while the propagating
solution is valid in the opposite, open limit, where the $\la_2$ is much
larger than the longitudinal dimension.

{\bf Theory}.\ -\
In what follows, we shall derive the general stationary solution of superfluid
hydrodynamics for the temperature $T$, the counterflow velocity
$w\eq\r_s(v_n-v_s)/\r_n$ and the $A$-$B$ interface velocity $\9u$, taking into
account the relaxation term $\r v_n/\tau$ in the momentum conservation,
and the altered connecting condition, Eq.~(\ref{Dp}). The solutions for $T$
and $w$ we search for are stationary in the rest frame of the interface,
with the $B$ phase at $x<0$ and the $A$ phase at $x>0$. Before the actual
calculation, however, we need to point out the following consideration that
significantly simplifies our task: As long as the radius of the experimental
cell is not too small, $R\agt 0.1$cm, the physics close to the interface,
especially the sq- and the diffusive lengths, are little changed from the
infinite geometry, and most of the results from Ref.~\cite{koliu93}
remain valid and need not be recalculated. More specifically, we have three
characteristic length scales:
(i) The microscopic width of the interface, too small to be considered in a
hydrodynamic approach except as the location of discontinuities.
(ii) The large decay length of second sound, Eq.~(\ref{la2}),
$\la_2\ap-\la_Rc_2/\9u$ for $\9u\ll c_2$ and
$\la_2\ap\la_R\9u/c_2$ for $\9u\gg c_2$, where the characteristic length
$\la_R\eq\tau c_2\r_n/\r_s$ is of order $0.1$cm for radii $R\sim 0.1$cm.
(iii) The mesoscopic decay lengths of various exponential functions.
Although individually describable in a hydrodynamic theory, they may also be
collectively accounted for by a total growth resistance of an effective
interface~\cite{koliu93} (that is wide enough to include these decays).
The decay lengths are the sq-length $\la_{sq}$ for $\9u\ll c_2$ and the
diffusive lengths $\la^D_T$, $\la^D_w$ for $\9u\gg c_2$.
In an infinite geometry, all three
are combinations of two characteristic lengths, both of order $10^{-2}$cm:
$\lambda_T\equiv k/(2c_2\rho T\6_T\sigma)$ and $\lambda_w\equiv [(4/3)\eta-\rho
(\zeta_1+\zeta_4)+\zeta_2+\rho^2\zeta_3]\rho_s/(2\rho\rho_n c_2)$. (The heat
conductance $k$ and the viscosities $\eta$, $\zeta_{1-4}$ are defined in the
usual way~\cite{vw}, neglecting anisotropy.) In a finite geometry, the
mesoscopic decay lengths are modified in first order of $\la_T/\la_R$,
$\la_w/\la_R$ as
\beq \la_{sq}^{A,B}=2(\la_T\la_w)^{1/2}\lb(1\mp{\la_T+\la_w\0 2(\la_T\la_w)
           ^{1/2}}{\9u\0 c_2}-{\la_T\0\la_R-\la_T}\rb) \label{lsq} \eeq
and
\beq \la^D_T=2\la_T{c_2\0\9u}\lb(1-{2\la_T\0\la_T-\la_w}{\la_T\0\la_R}
       {c_2^2\0\9u^2}\rb)\ ,\qd
     \la^D_w=2\la_w{c_2\0\9u}\lb(1+{2\la_w\0\la_T-\la_w}{\la_w\0\la_R}
       {c_2^2\0\9u^2}\rb)\ . \label{lTw} \eeq
The correction terms obviously may be safely
neglected for radii $R\agt 0.1$cm. For this experimentally relevant case
(the only one we study~\cite{Legg}), we may therefore take the effective
interface as calculated in Ref.~\cite{koliu93} for the infinite geometry,
especially the total growth resistance, and only consider the change further
away from the interface, on length scales of $\la_R$.
(Technically, this allows us to discard all dissipative terms in the
hydrodynamic equations except the new relaxation term $\r v_n/\tau$, cf. the
detailed discussion on the effective interface in Ref.~\cite{koliu93}.)

Linearizing the averaged hydrodynamic equations with respect to
$w\equiv \r_s(v_n-v_s)/\r$ and $T^{A,B}-T_i$, and retaining terms of
second order in $\9 u/c_2$ and $c_2/\9 u$ respectively, the stationary
solution for $\9 u\ll c_2$ is given by
\aea  \setcounter{equation}{0} \label{al5} \stepcounter{equation}
 T^B(x)&=&T_i+\de T^B\;,\ T^A(x)=T_i+\de T_2^A\,
                                    \exp\!\lb(-{\9u\0c_2}{x\0\la_R}\rb)\;, \\
 w^B(x)&=&0\;,\
 w^A(x)=\de w_2^A\,\exp\!\lb(-{\9u\0c_2}{x\0\la_R}\rb)\;,
\zea
and for $\9 u\gg c_2$ by
\aea  \setcounter{equation}{0} \label{al6} \stepcounter{equation}
 T^B(x)&=&T_i+\de T^B+\de T_2^B\,
           \exp\!\lb({c_2\0\9u}{x\0\la_R}\rb)\;,\ T^A(x)=T_i\;,  \\
 w^B(x)&=&\de w_2^B\,\exp\!\lb({c_2\0\9u}{x\0\la_R}\rb)\;,\ w^A(x)=0\;.
\zea
As discussed above, second-sound and the related energy transfer
now acquires a finite range. Its decay length
$\lambda_2=-\la_R(c_2^2-\dot u^2)/(c_2\dot u)$ is negative for
$\dot u<c_2$ and positive for $\dot u>c_2$. (Therefore, $\delta T_2^B$,
$\delta w_2^B=0$ in the first and $\delta T_2^A$, $\delta w_2^A=0$ in the
second case.) In a confined geometry, the amplitudes of the stationary
second-sound modes are related by
\beq \de w_2^{A,B}=\9u\,(\6_T\s/\s)\,\de T_2^{A,B} \eeq
(in contrast to the well-known relation
$\de w_2^{A,B}=\pm c_2\,(\6_T\s/\s)\,\de T_2^{A,B}$ for the infinite geometry).
The constant $\delta T^B$ accounts for the temperature rise
due to the latent heat and the interface dissipation, while $\delta T^A=0$
due to causality and the lack of any means for long range energy transfer.
The temperature fields (outside the effective interface), are plotted in
Fig.~1. and 2.

The interface velocity $\9 u$ and the independent amplitudes
of Eqs.~(\ref{al5}), (\ref{al6}) are to be determined from the general
connecting conditions (CCs) at the effective interface:
\bea \D Q=0\;,\qd\D g=0\;&,&\qd\Delta p=\Gamma\rho\dot u/R\;, \\
     \D(\mu+v_nv_s)=0\;&,&\qd R_s=\< f\>\,\D T+g\,\D\mu\;.
\eea
All quantities are defined in the rest frame of the interface and have been
averaged over the cross section of the experimental cell. The mass current
is given as $g=-\r\9u$~\cite{graliu90,koliu93}, $\<f\>$ denotes the average
$\2(f^B+f^A)$ of the entropy current across the interface.
Except the non-vanishing discontinuity of the pressure
$p$, all CCs are as derived in Ref.~\cite{koliu93}, though the subscript $e$
for ``effective'' has been eliminated from $\D$ and $\<\;\>$. From the
derivation there, it is also clear that $R_s$ is the total entropy production
within the effective interface. The entropy produced by the drag exerted by
the lateral walls at the rate of $\dot u\Delta p=\Gamma\rho\dot u^2/R$ is
therefore implicitly included in $R_s$. Subtracting this part, one obtains
the rate of entropy produced in an infinite geometry
\beq R_s^\infty=R_s-\G\r\9u^2/R\ , \label{Rs} \eeq
which was parametrized as $R_s^\infty=\<f\>^2/\k_e$ and $g^2/K_e$,
respectively for $\9u\ll c_2$ and $\9u\gg c_2$; cf.~\cite{graliu90}.
(Explicit expressions for the effective, R-independent Onsager coefficients
$\k_e$, $K_e$ can be found in Ref.~\cite{koliu93}.)
Since the term $\G\r\9u^2/R$ in Eq.~(\ref{Rs}) can be rewritten as
$\G\<f\>^2/(R\r\<\s\>^2)$ and $\G g^2/(R\r)$, we finally obtain
from $R_s\sim\<f\>^2$ and $g^2$ the CC:
%\FL
\begin{eqnarray}
  \langle f\rangle&=&\Bigl(\kappa_e^{-1}+\Gamma/(R\rho
      \langle\sigma\rangle^2)\Bigr)^{-1}\Delta T\ \ \ \ \ \ \ {\rm and} \\
  g&=&\Bigl(K_e^{-1}+\Gamma/(R\rho)\Bigr)^{-1}\bigl(\Delta\mu_o+
               \langle\sigma\rangle\,\Delta T\bigr)\ ,
\end{eqnarray}
respectively for $\dot u\ll c_2$ and $\dot u\gg c_2$. Here, $\mu_o$ denotes
the chemical potential for given pressure and temperature, in
a system with $v_n=v_s=0$.

Expanding the CCs to first order in the deviation from the initial
temperature $T_i$, we obtain a set of equations for the interface velocity
$\dot u$ and the amplitudes $\delta T^B$, $\delta T_2^A$ (or $\delta w_2^B$).
In the limit of $\dot u\ll c_2$, it is solved by
\aea  \setcounter{equation}{0} \label{res1} \stepcounter{equation}
   \de T^B&=&-\lb[{\D\s\0\6_T\s^B}+{\D\mu_o+\G\9u R^{-1}\0T\6_T\s^B}\rb]_i\ ,\\
 \de T_2^A&=&\de T^B-\lb[{1\0\s^A}\lb({(\D\s)^2\0\6_T\s^B}+\D\mu_o+\G\9u R^{-1}
                                       \rb)\rb]_i\ , \\
       \9u&=&-\lb({\r\<\s\>_i\0\k_e}+{2\G\0R}\rb)^{-1}\lb[{(\D\s)^2\0\6_T\s^B}
              +\D\mu_o\rb]_i\ ,
\zea
while the dependent counterflow amplitude is given as
$\de w_2^A=\9u[\6_T\s^A/\s^A]_i\,\de T_2^A$. Note the $R$-dependence of the
constant temperature $\delta T^B$, caused by the drag $\sim\G$.
Since $\dot u$ is taken to be positive here (melting the $A$ phase on the
right), the validity of Eqs.~(\ref{res1}) is confined to initial temperatures
satisfying $[|\D\mu_o|-(\D\s)^2/\6_T\s^B]_i\geq 0$, cf. the
discussion in the introduction.
For $\dot u\gg c_2$, expansion of the CCs around $T_i$ and
linearization with respect to $c_2/\dot u$, $\delta T^B$ and $\delta w_2^B$
lead to
\aea  \setcounter{equation}{0} \label{res2} \stepcounter{equation}
   \de T^B&=&-\lb[{\D\s\0\6_T\s^B}+{\D\mu_o+\G\9uR^{-1}\0T\6_T\s^B}\rb]_i\ ,\\
 \de w_2^B&=&-\lb[{\r_s\0\r_n}{\D\mu_o+\G\9uR^{-1}-\s^B\de T^B\0\9u}\rb]_i\ ,\\
       \9u&=&-\lb({\r\0K_e}+{2\G\0R}\rb)^{-1}\bl[\D\mu_o-\2\D\s\de T^B\br]_i\ ,
\zea
while the dependent temperature amplitude is given as
$\de T_2^B=(1/\9u)[\s^B/\6_T\s^B]_i\,\de w_2^B$.
Again, $\delta T^B$ contains an $R$-dependent contribution from the
latent heat and the interface entropy-production $R_s$; cf. Eq.~(\ref{res1}a).

{\bf Summary}.\ -\
The influence of finite geometries is considered, the interface velocity $\9u$
and the stationary temperature field are derived.
For both $\9u\ll c_2$ and $\9u\gg c_2$, the total growth resistance of the
transition is a sum of the open geometry Onsager-coefficient and a
``wall contribution'' $\sim\G/R$.
Hence $\9u$ (for given $T_i$) is  maximal for $R\to\infty$, as measured
\cite{boyd92}. The phenomenological parameter $\G$ depends on the cell's
inner surface, cf. Eq.~(\ref{Dp}). Two sets of $\9u$, obtained in tubes of
different radii $R$, should suffice to determine $\G$. This is not possible
with the published data~\cite{boyd92}, however, because the tubes used were
made of different materials, copper and epoxy, with presumably rather
different $\G$. Perhaps further experimentals will fill this gap.

Defining a total growth coefficient as
$K_{tot}(T_i/T_{AB}\,,\,H)\equiv(1/K_e+2\G/R\r)^{-1}$, we estimate
from the data of Ref.~\cite{boyd92} that
$K_{tot}(0.81\,,\,1.5$kOe)$=1.9\,{\rm g\,s\,cm^{-4}}\;,$
$K_{tot}(0.81\,,\,3.0$kOe)$=2.8\,{\rm g\,s\,cm^{-4}}\;,$
$K_{tot}(0.85\,,\,1.5$kOe)$=1.3\,{\rm g\,s\,cm^{-4}}\;,$
$K_{tot}(0.85\,,\,3.0$kOe)$=1.4\,{\rm g\,s\,cm^{-4}}\;;$
obviously, $K_{tot}$ increases with increasing magnetic field or
decreasing supercooling rate.

A paper containing the algebraic details of this letter and of
Ref.~\cite{koliu93} is being prepared.

{\bf Acknowledgment}.\ -\ One of the authors (P.K.) acknowledges financial
support of the Deutsche Forschungsgemeinschaft (DFG).

\tightenlines

\newpage
\vspace*{15mm}
\begin{center}
\setlength{\unitlength}{0.240900pt}
\ifx\plotpoint\undefined\newsavebox{\plotpoint}\fi
\sbox{\plotpoint}{\rule[-0.175pt]{0.350pt}{0.350pt}}%
\begin{picture}(1500,720)(0,0)
%\tenrm
\sbox{\plotpoint}{\rule[-0.175pt]{0.350pt}{0.350pt}}%
\put(264,158){\rule[-0.175pt]{282.335pt}{0.350pt}}
\put(850,158){\rule[-0.175pt]{0.350pt}{108.164pt}}
\put(1465,158){\makebox(0,0)[l]{$x$}}
\put(850,652){\makebox(0,0){$T(x)-T_i$}}
\put(498,248){\makebox(0,0){$^3He$-$B$}}
\put(1202,427){\makebox(0,0){$^3He$-$A$}}
\put(264,158){\vector(1,0){1172}}
\put(850,158){\vector(0,1){449}}
\put(264,374){\usebox{\plotpoint}}
\put(264,374){\rule[-0.175pt]{141.167pt}{0.350pt}}
\put(850,158){\rule[-0.175pt]{0.350pt}{52.034pt}}
\put(850,158){\rule[-0.175pt]{141.167pt}{0.350pt}}
\sbox{\plotpoint}{\rule[-0.350pt]{0.700pt}{0.700pt}}%
\put(264,158){\usebox{\plotpoint}}
\put(264,158){\rule[-0.350pt]{141.167pt}{0.700pt}}
\put(850,158){\rule[-0.350pt]{0.700pt}{86.483pt}}
\put(850,517){\usebox{\plotpoint}}
\put(851,515){\usebox{\plotpoint}}
\put(851,515){\usebox{\plotpoint}}
\put(852,514){\usebox{\plotpoint}}
\put(852,514){\usebox{\plotpoint}}
\put(853,513){\usebox{\plotpoint}}
\put(853,513){\usebox{\plotpoint}}
\put(854,512){\usebox{\plotpoint}}
\put(854,512){\usebox{\plotpoint}}
\put(855,511){\usebox{\plotpoint}}
\put(855,511){\usebox{\plotpoint}}
\put(856,509){\usebox{\plotpoint}}
\put(856,509){\usebox{\plotpoint}}
\put(857,508){\usebox{\plotpoint}}
\put(857,508){\usebox{\plotpoint}}
\put(858,507){\usebox{\plotpoint}}
\put(858,507){\usebox{\plotpoint}}
\put(859,506){\usebox{\plotpoint}}
\put(859,506){\usebox{\plotpoint}}
\put(860,505){\usebox{\plotpoint}}
\put(860,505){\usebox{\plotpoint}}
\put(861,503){\usebox{\plotpoint}}
\put(861,503){\usebox{\plotpoint}}
\put(862,502){\usebox{\plotpoint}}
\put(862,502){\usebox{\plotpoint}}
\put(863,501){\usebox{\plotpoint}}
\put(863,501){\usebox{\plotpoint}}
\put(864,500){\usebox{\plotpoint}}
\put(864,500){\usebox{\plotpoint}}
\put(865,499){\usebox{\plotpoint}}
\put(865,499){\usebox{\plotpoint}}
\put(866,498){\usebox{\plotpoint}}
\put(866,498){\usebox{\plotpoint}}
\put(867,496){\usebox{\plotpoint}}
\put(867,496){\usebox{\plotpoint}}
\put(868,495){\usebox{\plotpoint}}
\put(868,495){\usebox{\plotpoint}}
\put(869,494){\usebox{\plotpoint}}
\put(869,494){\usebox{\plotpoint}}
\put(870,493){\usebox{\plotpoint}}
\put(870,493){\usebox{\plotpoint}}
\put(871,492){\usebox{\plotpoint}}
\put(871,492){\usebox{\plotpoint}}
\put(872,491){\usebox{\plotpoint}}
\put(872,491){\usebox{\plotpoint}}
\put(873,490){\usebox{\plotpoint}}
\put(873,490){\usebox{\plotpoint}}
\put(874,488){\usebox{\plotpoint}}
\put(874,488){\usebox{\plotpoint}}
\put(875,487){\usebox{\plotpoint}}
\put(875,487){\usebox{\plotpoint}}
\put(876,486){\usebox{\plotpoint}}
\put(876,486){\usebox{\plotpoint}}
\put(877,485){\usebox{\plotpoint}}
\put(877,485){\usebox{\plotpoint}}
\put(878,484){\usebox{\plotpoint}}
\put(878,484){\usebox{\plotpoint}}
\put(879,483){\usebox{\plotpoint}}
\put(879,483){\usebox{\plotpoint}}
\put(880,482){\usebox{\plotpoint}}
\put(880,482){\usebox{\plotpoint}}
\put(881,481){\usebox{\plotpoint}}
\put(881,481){\usebox{\plotpoint}}
\put(882,480){\usebox{\plotpoint}}
\put(883,478){\usebox{\plotpoint}}
\put(883,478){\usebox{\plotpoint}}
\put(884,477){\usebox{\plotpoint}}
\put(884,477){\usebox{\plotpoint}}
\put(885,476){\usebox{\plotpoint}}
\put(885,476){\usebox{\plotpoint}}
\put(886,475){\usebox{\plotpoint}}
\put(886,475){\usebox{\plotpoint}}
\put(887,474){\usebox{\plotpoint}}
\put(887,474){\usebox{\plotpoint}}
\put(888,473){\usebox{\plotpoint}}
\put(888,473){\usebox{\plotpoint}}
\put(889,472){\usebox{\plotpoint}}
\put(889,472){\usebox{\plotpoint}}
\put(890,471){\usebox{\plotpoint}}
\put(890,471){\usebox{\plotpoint}}
\put(891,470){\usebox{\plotpoint}}
\put(891,470){\usebox{\plotpoint}}
\put(892,469){\usebox{\plotpoint}}
\put(892,469){\usebox{\plotpoint}}
\put(893,468){\usebox{\plotpoint}}
\put(893,468){\usebox{\plotpoint}}
\put(894,467){\usebox{\plotpoint}}
\put(894,467){\usebox{\plotpoint}}
\put(895,466){\usebox{\plotpoint}}
\put(896,465){\usebox{\plotpoint}}
\put(897,464){\usebox{\plotpoint}}
\put(898,462){\usebox{\plotpoint}}
\put(898,462){\usebox{\plotpoint}}
\put(899,461){\usebox{\plotpoint}}
\put(899,461){\usebox{\plotpoint}}
\put(900,460){\usebox{\plotpoint}}
\put(900,460){\usebox{\plotpoint}}
\put(901,459){\usebox{\plotpoint}}
\put(901,459){\usebox{\plotpoint}}
\put(902,458){\usebox{\plotpoint}}
\put(902,458){\usebox{\plotpoint}}
\put(903,457){\usebox{\plotpoint}}
\put(903,457){\usebox{\plotpoint}}
\put(904,456){\usebox{\plotpoint}}
\put(904,456){\usebox{\plotpoint}}
\put(905,455){\usebox{\plotpoint}}
\put(905,455){\usebox{\plotpoint}}
\put(906,454){\usebox{\plotpoint}}
\put(906,454){\usebox{\plotpoint}}
\put(907,453){\usebox{\plotpoint}}
\put(907,453){\usebox{\plotpoint}}
\put(908,452){\usebox{\plotpoint}}
\put(908,452){\usebox{\plotpoint}}
\put(909,451){\usebox{\plotpoint}}
\put(909,451){\usebox{\plotpoint}}
\put(910,450){\usebox{\plotpoint}}
\put(910,450){\usebox{\plotpoint}}
\put(911,449){\usebox{\plotpoint}}
\put(911,449){\usebox{\plotpoint}}
\put(912,448){\usebox{\plotpoint}}
\put(912,448){\usebox{\plotpoint}}
\put(913,447){\usebox{\plotpoint}}
\put(913,447){\usebox{\plotpoint}}
\put(914,446){\usebox{\plotpoint}}
\put(914,446){\usebox{\plotpoint}}
\put(915,445){\usebox{\plotpoint}}
\put(915,445){\usebox{\plotpoint}}
\put(916,444){\usebox{\plotpoint}}
\put(916,444){\usebox{\plotpoint}}
\put(917,443){\usebox{\plotpoint}}
\put(917,443){\usebox{\plotpoint}}
\put(918,442){\usebox{\plotpoint}}
\put(918,442){\usebox{\plotpoint}}
\put(919,441){\usebox{\plotpoint}}
\put(919,441){\usebox{\plotpoint}}
\put(920,440){\usebox{\plotpoint}}
\put(920,440){\usebox{\plotpoint}}
\put(921,439){\usebox{\plotpoint}}
\put(921,439){\usebox{\plotpoint}}
\put(922,438){\usebox{\plotpoint}}
\put(922,438){\usebox{\plotpoint}}
\put(924,437){\usebox{\plotpoint}}
\put(925,436){\usebox{\plotpoint}}
\put(926,435){\usebox{\plotpoint}}
\put(926,435){\usebox{\plotpoint}}
\put(927,434){\usebox{\plotpoint}}
\put(927,434){\usebox{\plotpoint}}
\put(928,433){\usebox{\plotpoint}}
\put(928,433){\usebox{\plotpoint}}
\put(929,432){\usebox{\plotpoint}}
\put(929,432){\usebox{\plotpoint}}
\put(930,431){\usebox{\plotpoint}}
\put(930,431){\usebox{\plotpoint}}
\put(931,430){\usebox{\plotpoint}}
\put(931,430){\usebox{\plotpoint}}
\put(932,429){\usebox{\plotpoint}}
\put(932,429){\usebox{\plotpoint}}
\put(933,428){\usebox{\plotpoint}}
\put(933,428){\usebox{\plotpoint}}
\put(934,427){\usebox{\plotpoint}}
\put(934,427){\usebox{\plotpoint}}
\put(935,426){\usebox{\plotpoint}}
\put(935,426){\usebox{\plotpoint}}
\put(936,425){\usebox{\plotpoint}}
\put(936,425){\usebox{\plotpoint}}
\put(937,424){\usebox{\plotpoint}}
\put(937,424){\usebox{\plotpoint}}
\put(939,423){\usebox{\plotpoint}}
\put(939,423){\usebox{\plotpoint}}
\put(940,422){\usebox{\plotpoint}}
\put(940,422){\usebox{\plotpoint}}
\put(941,421){\usebox{\plotpoint}}
\put(941,421){\usebox{\plotpoint}}
\put(942,420){\usebox{\plotpoint}}
\put(942,420){\usebox{\plotpoint}}
\put(943,419){\usebox{\plotpoint}}
\put(943,419){\usebox{\plotpoint}}
\put(944,418){\usebox{\plotpoint}}
\put(944,418){\usebox{\plotpoint}}
\put(945,417){\usebox{\plotpoint}}
\put(945,417){\usebox{\plotpoint}}
\put(946,416){\usebox{\plotpoint}}
\put(946,416){\usebox{\plotpoint}}
\put(948,415){\usebox{\plotpoint}}
\put(949,414){\usebox{\plotpoint}}
\put(949,414){\usebox{\plotpoint}}
\put(950,413){\usebox{\plotpoint}}
\put(950,413){\usebox{\plotpoint}}
\put(951,412){\usebox{\plotpoint}}
\put(951,412){\usebox{\plotpoint}}
\put(952,411){\usebox{\plotpoint}}
\put(952,411){\usebox{\plotpoint}}
\put(953,410){\usebox{\plotpoint}}
\put(953,410){\usebox{\plotpoint}}
\put(954,409){\usebox{\plotpoint}}
\put(954,409){\usebox{\plotpoint}}
\put(956,408){\usebox{\plotpoint}}
\put(956,408){\usebox{\plotpoint}}
\put(957,407){\usebox{\plotpoint}}
\put(957,407){\usebox{\plotpoint}}
\put(958,406){\usebox{\plotpoint}}
\put(958,406){\usebox{\plotpoint}}
\put(959,405){\usebox{\plotpoint}}
\put(959,405){\usebox{\plotpoint}}
\put(960,404){\usebox{\plotpoint}}
\put(960,404){\usebox{\plotpoint}}
\put(962,403){\usebox{\plotpoint}}
\put(963,402){\usebox{\plotpoint}}
\put(963,402){\usebox{\plotpoint}}
\put(964,401){\usebox{\plotpoint}}
\put(964,401){\usebox{\plotpoint}}
\put(965,400){\usebox{\plotpoint}}
\put(965,400){\usebox{\plotpoint}}
\put(966,399){\usebox{\plotpoint}}
\put(966,399){\usebox{\plotpoint}}
\put(968,398){\usebox{\plotpoint}}
\put(968,398){\usebox{\plotpoint}}
\put(969,397){\usebox{\plotpoint}}
\put(969,397){\usebox{\plotpoint}}
\put(970,396){\usebox{\plotpoint}}
\put(970,396){\usebox{\plotpoint}}
\put(971,395){\usebox{\plotpoint}}
\put(971,395){\usebox{\plotpoint}}
\put(973,394){\usebox{\plotpoint}}
\put(974,393){\usebox{\plotpoint}}
\put(974,393){\usebox{\plotpoint}}
\put(975,392){\usebox{\plotpoint}}
\put(975,392){\usebox{\plotpoint}}
\put(976,391){\usebox{\plotpoint}}
\put(976,391){\usebox{\plotpoint}}
\put(977,390){\usebox{\plotpoint}}
\put(977,390){\usebox{\plotpoint}}
\put(979,389){\usebox{\plotpoint}}
\put(979,389){\usebox{\plotpoint}}
\put(980,388){\usebox{\plotpoint}}
\put(980,388){\usebox{\plotpoint}}
\put(981,387){\usebox{\plotpoint}}
\put(981,387){\usebox{\plotpoint}}
\put(983,386){\usebox{\plotpoint}}
\put(983,386){\usebox{\plotpoint}}
\put(984,385){\usebox{\plotpoint}}
\put(984,385){\usebox{\plotpoint}}
\put(985,384){\usebox{\plotpoint}}
\put(985,384){\usebox{\plotpoint}}
\put(986,383){\usebox{\plotpoint}}
\put(986,383){\usebox{\plotpoint}}
\put(988,382){\usebox{\plotpoint}}
\put(988,382){\usebox{\plotpoint}}
\put(989,381){\usebox{\plotpoint}}
\put(989,381){\usebox{\plotpoint}}
\put(990,380){\usebox{\plotpoint}}
\put(990,380){\usebox{\plotpoint}}
\put(992,379){\usebox{\plotpoint}}
\put(992,379){\usebox{\plotpoint}}
\put(993,378){\usebox{\plotpoint}}
\put(993,378){\usebox{\plotpoint}}
\put(994,377){\usebox{\plotpoint}}
\put(994,377){\usebox{\plotpoint}}
\put(996,376){\usebox{\plotpoint}}
\put(996,376){\usebox{\plotpoint}}
\put(997,375){\usebox{\plotpoint}}
\put(997,375){\usebox{\plotpoint}}
\put(998,374){\usebox{\plotpoint}}
\put(998,374){\usebox{\plotpoint}}
\put(1000,373){\usebox{\plotpoint}}
\put(1000,373){\usebox{\plotpoint}}
\put(1001,372){\usebox{\plotpoint}}
\put(1001,372){\usebox{\plotpoint}}
\put(1002,371){\usebox{\plotpoint}}
\put(1002,371){\usebox{\plotpoint}}
\put(1004,370){\usebox{\plotpoint}}
\put(1004,370){\usebox{\plotpoint}}
\put(1005,369){\usebox{\plotpoint}}
\put(1005,369){\usebox{\plotpoint}}
\put(1007,368){\usebox{\plotpoint}}
\put(1007,368){\usebox{\plotpoint}}
\put(1008,367){\usebox{\plotpoint}}
\put(1008,367){\usebox{\plotpoint}}
\put(1009,366){\usebox{\plotpoint}}
\put(1009,366){\usebox{\plotpoint}}
\put(1011,365){\usebox{\plotpoint}}
\put(1011,365){\usebox{\plotpoint}}
\put(1012,364){\usebox{\plotpoint}}
\put(1012,364){\usebox{\plotpoint}}
\put(1014,363){\usebox{\plotpoint}}
\put(1014,363){\usebox{\plotpoint}}
\put(1015,362){\usebox{\plotpoint}}
\put(1015,362){\usebox{\plotpoint}}
\put(1016,361){\usebox{\plotpoint}}
\put(1016,361){\usebox{\plotpoint}}
\put(1018,360){\usebox{\plotpoint}}
\put(1018,360){\usebox{\plotpoint}}
\put(1019,359){\usebox{\plotpoint}}
\put(1019,359){\usebox{\plotpoint}}
\put(1021,358){\usebox{\plotpoint}}
\put(1021,358){\usebox{\plotpoint}}
\put(1022,357){\usebox{\plotpoint}}
\put(1022,357){\usebox{\plotpoint}}
\put(1024,356){\usebox{\plotpoint}}
\put(1024,356){\usebox{\plotpoint}}
\put(1025,355){\usebox{\plotpoint}}
\put(1025,355){\usebox{\plotpoint}}
\put(1027,354){\usebox{\plotpoint}}
\put(1027,354){\usebox{\plotpoint}}
\put(1028,353){\usebox{\plotpoint}}
\put(1028,353){\usebox{\plotpoint}}
\put(1030,352){\usebox{\plotpoint}}
\put(1030,352){\usebox{\plotpoint}}
\put(1031,351){\usebox{\plotpoint}}
\put(1031,351){\usebox{\plotpoint}}
\put(1033,350){\usebox{\plotpoint}}
\put(1033,350){\usebox{\plotpoint}}
\put(1034,349){\usebox{\plotpoint}}
\put(1034,349){\usebox{\plotpoint}}
\put(1036,348){\usebox{\plotpoint}}
\put(1036,348){\usebox{\plotpoint}}
\put(1037,347){\usebox{\plotpoint}}
\put(1037,347){\usebox{\plotpoint}}
\put(1039,346){\usebox{\plotpoint}}
\put(1039,346){\usebox{\plotpoint}}
\put(1041,345){\usebox{\plotpoint}}
\put(1042,344){\usebox{\plotpoint}}
\put(1042,344){\usebox{\plotpoint}}
\put(1044,343){\usebox{\plotpoint}}
\put(1044,343){\usebox{\plotpoint}}
\put(1045,342){\usebox{\plotpoint}}
\put(1045,342){\usebox{\plotpoint}}
\put(1047,341){\usebox{\plotpoint}}
\put(1047,341){\usebox{\plotpoint}}
\put(1048,340){\usebox{\plotpoint}}
\put(1048,340){\usebox{\plotpoint}}
\put(1050,339){\usebox{\plotpoint}}
\put(1050,339){\usebox{\plotpoint}}
\put(1052,338){\usebox{\plotpoint}}
\put(1052,338){\usebox{\plotpoint}}
\put(1053,337){\usebox{\plotpoint}}
\put(1053,337){\usebox{\plotpoint}}
\put(1055,336){\usebox{\plotpoint}}
\put(1055,336){\usebox{\plotpoint}}
\put(1057,335){\usebox{\plotpoint}}
\put(1058,334){\usebox{\plotpoint}}
\put(1058,334){\usebox{\plotpoint}}
\put(1060,333){\usebox{\plotpoint}}
\put(1060,333){\usebox{\plotpoint}}
\put(1062,332){\usebox{\plotpoint}}
\put(1062,332){\usebox{\plotpoint}}
\put(1063,331){\usebox{\plotpoint}}
\put(1063,331){\usebox{\plotpoint}}
\put(1065,330){\usebox{\plotpoint}}
\put(1065,330){\usebox{\plotpoint}}
\put(1067,329){\usebox{\plotpoint}}
\put(1067,329){\usebox{\plotpoint}}
\put(1068,328){\usebox{\plotpoint}}
\put(1068,328){\usebox{\plotpoint}}
\put(1070,327){\usebox{\plotpoint}}
\put(1070,327){\usebox{\plotpoint}}
\put(1072,326){\usebox{\plotpoint}}
\put(1072,326){\usebox{\plotpoint}}
\put(1074,325){\usebox{\plotpoint}}
\put(1075,324){\usebox{\plotpoint}}
\put(1075,324){\usebox{\plotpoint}}
\put(1077,323){\usebox{\plotpoint}}
\put(1077,323){\usebox{\plotpoint}}
\put(1079,322){\usebox{\plotpoint}}
\put(1079,322){\usebox{\plotpoint}}
\put(1081,321){\usebox{\plotpoint}}
\put(1082,320){\usebox{\plotpoint}}
\put(1082,320){\usebox{\plotpoint}}
\put(1084,319){\usebox{\plotpoint}}
\put(1084,319){\usebox{\plotpoint}}
\put(1086,318){\usebox{\plotpoint}}
\put(1086,318){\usebox{\plotpoint}}
\put(1088,317){\usebox{\plotpoint}}
\put(1088,317){\usebox{\plotpoint}}
\put(1090,316){\usebox{\plotpoint}}
\put(1090,316){\usebox{\plotpoint}}
\put(1092,315){\usebox{\plotpoint}}
\put(1092,315){\usebox{\plotpoint}}
\put(1094,314){\usebox{\plotpoint}}
\put(1095,313){\usebox{\plotpoint}}
\put(1095,313){\usebox{\plotpoint}}
\put(1097,312){\usebox{\plotpoint}}
\put(1097,312){\usebox{\plotpoint}}
\put(1099,311){\usebox{\plotpoint}}
\put(1099,311){\usebox{\plotpoint}}
\put(1101,310){\usebox{\plotpoint}}
\put(1101,310){\usebox{\plotpoint}}
\put(1103,309){\usebox{\plotpoint}}
\put(1103,309){\usebox{\plotpoint}}
\put(1105,308){\usebox{\plotpoint}}
\put(1105,308){\usebox{\plotpoint}}
\put(1107,307){\usebox{\plotpoint}}
\put(1107,307){\usebox{\plotpoint}}
\put(1109,306){\usebox{\plotpoint}}
\put(1109,306){\usebox{\plotpoint}}
\put(1111,305){\usebox{\plotpoint}}
\put(1111,305){\usebox{\plotpoint}}
\put(1113,304){\usebox{\plotpoint}}
\put(1113,304){\usebox{\plotpoint}}
\put(1115,303){\usebox{\plotpoint}}
\put(1115,303){\usebox{\plotpoint}}
\put(1117,302){\usebox{\plotpoint}}
\put(1117,302){\usebox{\plotpoint}}
\put(1119,301){\usebox{\plotpoint}}
\put(1119,301){\usebox{\plotpoint}}
\put(1121,300){\usebox{\plotpoint}}
\put(1121,300){\usebox{\plotpoint}}
\put(1123,299){\usebox{\plotpoint}}
\put(1123,299){\usebox{\plotpoint}}
\put(1125,298){\usebox{\plotpoint}}
\put(1125,298){\usebox{\plotpoint}}
\put(1127,297){\usebox{\plotpoint}}
\put(1127,297){\usebox{\plotpoint}}
\put(1129,296){\usebox{\plotpoint}}
\put(1129,296){\usebox{\plotpoint}}
\put(1131,295){\usebox{\plotpoint}}
\put(1131,295){\rule[-0.350pt]{0.723pt}{0.700pt}}
\put(1134,294){\usebox{\plotpoint}}
\put(1136,293){\usebox{\plotpoint}}
\put(1136,293){\usebox{\plotpoint}}
\put(1138,292){\usebox{\plotpoint}}
\put(1138,292){\usebox{\plotpoint}}
\put(1140,291){\usebox{\plotpoint}}
\put(1140,291){\usebox{\plotpoint}}
\put(1142,290){\usebox{\plotpoint}}
\put(1142,290){\usebox{\plotpoint}}
\put(1144,289){\usebox{\plotpoint}}
\put(1144,289){\rule[-0.350pt]{0.723pt}{0.700pt}}
\put(1147,288){\usebox{\plotpoint}}
\put(1147,288){\usebox{\plotpoint}}
\put(1149,287){\usebox{\plotpoint}}
\put(1149,287){\usebox{\plotpoint}}
\put(1151,286){\usebox{\plotpoint}}
\put(1151,286){\rule[-0.350pt]{0.723pt}{0.700pt}}
\put(1154,285){\usebox{\plotpoint}}
\put(1156,284){\usebox{\plotpoint}}
\put(1156,284){\usebox{\plotpoint}}
\put(1158,283){\usebox{\plotpoint}}
\put(1158,283){\rule[-0.350pt]{0.723pt}{0.700pt}}
\put(1161,282){\usebox{\plotpoint}}
\put(1163,281){\usebox{\plotpoint}}
\put(1163,281){\usebox{\plotpoint}}
\put(1165,280){\usebox{\plotpoint}}
\put(1165,280){\rule[-0.350pt]{0.723pt}{0.700pt}}
\put(1168,279){\usebox{\plotpoint}}
\put(1168,279){\usebox{\plotpoint}}
\put(1170,278){\usebox{\plotpoint}}
\put(1170,278){\rule[-0.350pt]{0.723pt}{0.700pt}}
\put(1173,277){\usebox{\plotpoint}}
\put(1175,276){\usebox{\plotpoint}}
\put(1175,276){\usebox{\plotpoint}}
\put(1177,275){\usebox{\plotpoint}}
\put(1177,275){\rule[-0.350pt]{0.723pt}{0.700pt}}
\put(1180,274){\usebox{\plotpoint}}
\put(1180,274){\usebox{\plotpoint}}
\put(1182,273){\usebox{\plotpoint}}
\put(1182,273){\rule[-0.350pt]{0.723pt}{0.700pt}}
\put(1185,272){\usebox{\plotpoint}}
\put(1185,272){\rule[-0.350pt]{0.723pt}{0.700pt}}
\put(1188,271){\usebox{\plotpoint}}
\put(1188,271){\usebox{\plotpoint}}
\put(1190,270){\usebox{\plotpoint}}
\put(1190,270){\rule[-0.350pt]{0.723pt}{0.700pt}}
\put(1193,269){\usebox{\plotpoint}}
\put(1193,269){\usebox{\plotpoint}}
\put(1195,268){\usebox{\plotpoint}}
\put(1195,268){\rule[-0.350pt]{0.723pt}{0.700pt}}
\put(1198,267){\usebox{\plotpoint}}
\put(1198,267){\rule[-0.350pt]{0.723pt}{0.700pt}}
\put(1201,266){\usebox{\plotpoint}}
\put(1201,266){\rule[-0.350pt]{0.723pt}{0.700pt}}
\put(1204,265){\usebox{\plotpoint}}
\put(1206,264){\usebox{\plotpoint}}
\put(1206,264){\rule[-0.350pt]{0.723pt}{0.700pt}}
\put(1209,263){\usebox{\plotpoint}}
\put(1209,263){\rule[-0.350pt]{0.723pt}{0.700pt}}
\put(1212,262){\usebox{\plotpoint}}
\put(1212,262){\rule[-0.350pt]{0.723pt}{0.700pt}}
\put(1215,261){\rule[-0.350pt]{0.723pt}{0.700pt}}
\put(1218,260){\usebox{\plotpoint}}
\put(1220,259){\usebox{\plotpoint}}
\put(1220,259){\rule[-0.350pt]{0.723pt}{0.700pt}}
\put(1223,258){\usebox{\plotpoint}}
\put(1223,258){\rule[-0.350pt]{0.723pt}{0.700pt}}
\put(1226,257){\usebox{\plotpoint}}
\put(1226,257){\rule[-0.350pt]{0.723pt}{0.700pt}}
\put(1229,256){\usebox{\plotpoint}}
\put(1229,256){\rule[-0.350pt]{0.723pt}{0.700pt}}
\put(1232,255){\usebox{\plotpoint}}
\put(1232,255){\rule[-0.350pt]{0.723pt}{0.700pt}}
\put(1235,254){\usebox{\plotpoint}}
\put(1235,254){\rule[-0.350pt]{0.723pt}{0.700pt}}
\put(1238,253){\usebox{\plotpoint}}
\put(1238,253){\rule[-0.350pt]{0.723pt}{0.700pt}}
\put(1241,252){\usebox{\plotpoint}}
\put(1241,252){\rule[-0.350pt]{0.723pt}{0.700pt}}
\put(1244,251){\usebox{\plotpoint}}
\put(1244,251){\rule[-0.350pt]{0.964pt}{0.700pt}}
\put(1248,250){\rule[-0.350pt]{0.723pt}{0.700pt}}
\put(1251,249){\usebox{\plotpoint}}
\put(1251,249){\rule[-0.350pt]{0.723pt}{0.700pt}}
\put(1254,248){\usebox{\plotpoint}}
\put(1254,248){\rule[-0.350pt]{0.723pt}{0.700pt}}
\put(1257,247){\usebox{\plotpoint}}
\put(1257,247){\rule[-0.350pt]{0.964pt}{0.700pt}}
\put(1261,246){\rule[-0.350pt]{0.723pt}{0.700pt}}
\put(1264,245){\usebox{\plotpoint}}
\put(1264,245){\rule[-0.350pt]{0.723pt}{0.700pt}}
\put(1267,244){\usebox{\plotpoint}}
\put(1267,244){\rule[-0.350pt]{0.964pt}{0.700pt}}
\put(1271,243){\rule[-0.350pt]{0.723pt}{0.700pt}}
\put(1274,242){\usebox{\plotpoint}}
\put(1274,242){\rule[-0.350pt]{0.964pt}{0.700pt}}
\put(1278,241){\rule[-0.350pt]{0.723pt}{0.700pt}}
\put(1281,240){\usebox{\plotpoint}}
\put(1281,240){\rule[-0.350pt]{0.964pt}{0.700pt}}
\put(1285,239){\usebox{\plotpoint}}
\put(1285,239){\rule[-0.350pt]{0.723pt}{0.700pt}}
\put(1288,238){\usebox{\plotpoint}}
\put(1288,238){\rule[-0.350pt]{0.964pt}{0.700pt}}
\put(1292,237){\usebox{\plotpoint}}
\put(1292,237){\rule[-0.350pt]{0.964pt}{0.700pt}}
\put(1296,236){\usebox{\plotpoint}}
\put(1296,236){\rule[-0.350pt]{0.723pt}{0.700pt}}
\put(1299,235){\usebox{\plotpoint}}
\put(1299,235){\rule[-0.350pt]{0.964pt}{0.700pt}}
\put(1303,234){\usebox{\plotpoint}}
\put(1303,234){\rule[-0.350pt]{0.964pt}{0.700pt}}
\put(1307,233){\usebox{\plotpoint}}
\put(1307,233){\rule[-0.350pt]{0.964pt}{0.700pt}}
\put(1311,232){\usebox{\plotpoint}}
\put(1311,232){\rule[-0.350pt]{0.964pt}{0.700pt}}
\put(1315,231){\usebox{\plotpoint}}
\put(1315,231){\rule[-0.350pt]{0.964pt}{0.700pt}}
\put(1319,230){\usebox{\plotpoint}}
\put(1319,230){\rule[-0.350pt]{0.964pt}{0.700pt}}
\put(1323,229){\usebox{\plotpoint}}
\put(1323,229){\rule[-0.350pt]{0.964pt}{0.700pt}}
\put(1327,228){\usebox{\plotpoint}}
\put(1327,228){\rule[-0.350pt]{0.964pt}{0.700pt}}
\put(1331,227){\usebox{\plotpoint}}
\put(1331,227){\rule[-0.350pt]{1.204pt}{0.700pt}}
\put(1336,226){\rule[-0.350pt]{0.964pt}{0.700pt}}
\put(1340,225){\usebox{\plotpoint}}
\put(1340,225){\rule[-0.350pt]{0.964pt}{0.700pt}}
\put(1344,224){\usebox{\plotpoint}}
\put(1344,224){\rule[-0.350pt]{1.204pt}{0.700pt}}
\put(1349,223){\usebox{\plotpoint}}
\put(1349,223){\rule[-0.350pt]{0.964pt}{0.700pt}}
\put(1353,222){\usebox{\plotpoint}}
\put(1353,222){\rule[-0.350pt]{1.204pt}{0.700pt}}
\put(1358,221){\usebox{\plotpoint}}
\put(1358,221){\rule[-0.350pt]{0.964pt}{0.700pt}}
\put(1362,220){\usebox{\plotpoint}}
\put(1362,220){\rule[-0.350pt]{1.204pt}{0.700pt}}
\put(1367,219){\usebox{\plotpoint}}
\put(1367,219){\rule[-0.350pt]{1.204pt}{0.700pt}}
\put(1372,218){\usebox{\plotpoint}}
\put(1372,218){\rule[-0.350pt]{1.204pt}{0.700pt}}
\put(1377,217){\usebox{\plotpoint}}
\put(1377,217){\rule[-0.350pt]{1.204pt}{0.700pt}}
\put(1382,216){\usebox{\plotpoint}}
\put(1382,216){\rule[-0.350pt]{1.204pt}{0.700pt}}
\put(1387,215){\usebox{\plotpoint}}
\put(1387,215){\rule[-0.350pt]{1.204pt}{0.700pt}}
\put(1392,214){\usebox{\plotpoint}}
\put(1392,214){\rule[-0.350pt]{1.204pt}{0.700pt}}
\put(1397,213){\usebox{\plotpoint}}
\put(1397,213){\rule[-0.350pt]{1.445pt}{0.700pt}}
\put(1403,212){\rule[-0.350pt]{1.204pt}{0.700pt}}
\put(1408,211){\usebox{\plotpoint}}
\put(1408,211){\rule[-0.350pt]{1.204pt}{0.700pt}}
\put(1413,210){\usebox{\plotpoint}}
\put(1413,210){\rule[-0.350pt]{1.445pt}{0.700pt}}
\put(1419,209){\usebox{\plotpoint}}
\put(1419,209){\rule[-0.350pt]{1.445pt}{0.700pt}}
\put(1425,208){\usebox{\plotpoint}}
\put(1425,208){\rule[-0.350pt]{1.445pt}{0.700pt}}
\put(1431,207){\usebox{\plotpoint}}
\put(1431,207){\rule[-0.350pt]{1.204pt}{0.700pt}}
\end{picture}\end{center}
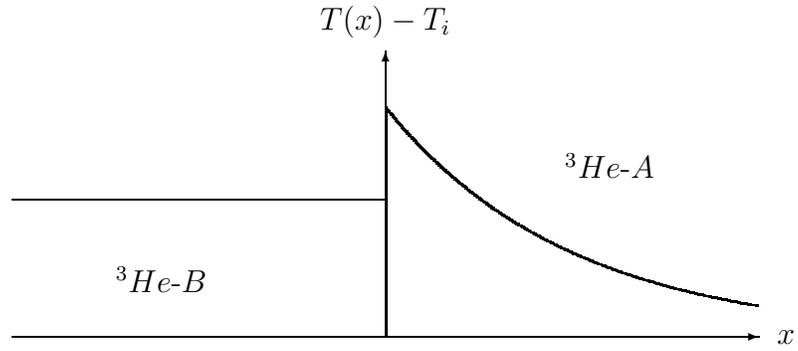
FIG.~1.\ \ The temperature field for $\dot u\ll c_2$, as in Eq.~(6a).
\vspace{20mm}
\begin{center}
\setlength{\unitlength}{0.240900pt}
\ifx\plotpoint\undefined\newsavebox{\plotpoint}\fi
\sbox{\plotpoint}{\rule[-0.175pt]{0.350pt}{0.350pt}}%
\begin{picture}(1500,720)(0,0)
%\tenrm
\sbox{\plotpoint}{\rule[-0.175pt]{0.350pt}{0.350pt}}%
\put(264,158){\rule[-0.175pt]{282.335pt}{0.350pt}}
\put(850,158){\rule[-0.175pt]{0.350pt}{108.164pt}}
\put(1465,158){\makebox(0,0)[l]{$x$}}
\put(850,652){\makebox(0,0){$T(x)-T_i$}}
\put(528,293){\makebox(0,0){$^3He$-$B$}}
\put(1172,293){\makebox(0,0){$^3He$-$A$}}
\put(264,158){\vector(1,0){1172}}
\put(850,158){\vector(0,1){449}}
\put(264,478){\usebox{\plotpoint}}
\put(264,478){\rule[-0.175pt]{1.445pt}{0.350pt}}
\put(270,478){\usebox{\plotpoint}}
\put(270,479){\rule[-0.175pt]{10.600pt}{0.350pt}}
\put(314,479){\usebox{\plotpoint}}
\put(314,480){\rule[-0.175pt]{9.154pt}{0.350pt}}
\put(352,480){\usebox{\plotpoint}}
\put(352,481){\rule[-0.175pt]{8.191pt}{0.350pt}}
\put(386,482){\rule[-0.175pt]{7.227pt}{0.350pt}}
\put(416,482){\usebox{\plotpoint}}
\put(416,483){\rule[-0.175pt]{6.504pt}{0.350pt}}
\put(443,483){\usebox{\plotpoint}}
\put(443,484){\rule[-0.175pt]{6.022pt}{0.350pt}}
\put(468,484){\usebox{\plotpoint}}
\put(468,485){\rule[-0.175pt]{5.541pt}{0.350pt}}
\put(491,485){\usebox{\plotpoint}}
\put(491,486){\rule[-0.175pt]{5.300pt}{0.350pt}}
\put(513,486){\usebox{\plotpoint}}
\put(513,487){\rule[-0.175pt]{4.818pt}{0.350pt}}
\put(533,487){\usebox{\plotpoint}}
\put(533,488){\rule[-0.175pt]{4.336pt}{0.350pt}}
\put(551,488){\usebox{\plotpoint}}
\put(551,489){\rule[-0.175pt]{4.336pt}{0.350pt}}
\put(569,489){\usebox{\plotpoint}}
\put(569,490){\rule[-0.175pt]{4.095pt}{0.350pt}}
\put(586,491){\rule[-0.175pt]{3.613pt}{0.350pt}}
\put(601,491){\usebox{\plotpoint}}
\put(601,492){\rule[-0.175pt]{3.613pt}{0.350pt}}
\put(616,492){\usebox{\plotpoint}}
\put(616,493){\rule[-0.175pt]{3.373pt}{0.350pt}}
\put(630,493){\usebox{\plotpoint}}
\put(630,494){\rule[-0.175pt]{3.373pt}{0.350pt}}
\put(644,494){\usebox{\plotpoint}}
\put(644,495){\rule[-0.175pt]{3.132pt}{0.350pt}}
\put(657,496){\rule[-0.175pt]{2.891pt}{0.350pt}}
\put(669,496){\usebox{\plotpoint}}
\put(669,497){\rule[-0.175pt]{2.891pt}{0.350pt}}
\put(681,497){\usebox{\plotpoint}}
\put(681,498){\rule[-0.175pt]{2.650pt}{0.350pt}}
\put(692,498){\usebox{\plotpoint}}
\put(692,499){\rule[-0.175pt]{2.650pt}{0.350pt}}
\put(703,499){\usebox{\plotpoint}}
\put(703,500){\rule[-0.175pt]{2.650pt}{0.350pt}}
\put(714,500){\usebox{\plotpoint}}
\put(714,501){\rule[-0.175pt]{2.409pt}{0.350pt}}
\put(724,501){\usebox{\plotpoint}}
\put(724,502){\rule[-0.175pt]{2.409pt}{0.350pt}}
\put(734,502){\usebox{\plotpoint}}
\put(734,503){\rule[-0.175pt]{2.168pt}{0.350pt}}
\put(743,503){\usebox{\plotpoint}}
\put(743,504){\rule[-0.175pt]{2.409pt}{0.350pt}}
\put(753,504){\usebox{\plotpoint}}
\put(753,505){\rule[-0.175pt]{2.168pt}{0.350pt}}
\put(762,506){\rule[-0.175pt]{1.927pt}{0.350pt}}
\put(770,506){\usebox{\plotpoint}}
\put(770,507){\rule[-0.175pt]{2.168pt}{0.350pt}}
\put(779,507){\usebox{\plotpoint}}
\put(779,508){\rule[-0.175pt]{1.927pt}{0.350pt}}
\put(787,508){\usebox{\plotpoint}}
\put(787,509){\rule[-0.175pt]{1.927pt}{0.350pt}}
\put(795,509){\usebox{\plotpoint}}
\put(795,510){\rule[-0.175pt]{1.927pt}{0.350pt}}
\put(803,510){\usebox{\plotpoint}}
\put(803,511){\rule[-0.175pt]{1.686pt}{0.350pt}}
\put(810,511){\usebox{\plotpoint}}
\put(810,512){\rule[-0.175pt]{1.927pt}{0.350pt}}
\put(818,512){\usebox{\plotpoint}}
\put(818,513){\rule[-0.175pt]{1.686pt}{0.350pt}}
\put(825,513){\usebox{\plotpoint}}
\put(825,514){\rule[-0.175pt]{1.686pt}{0.350pt}}
\put(832,514){\usebox{\plotpoint}}
\put(832,515){\rule[-0.175pt]{1.686pt}{0.350pt}}
\put(839,515){\usebox{\plotpoint}}
\put(839,516){\rule[-0.175pt]{1.445pt}{0.350pt}}
\put(845,516){\usebox{\plotpoint}}
\put(845,517){\rule[-0.175pt]{1.204pt}{0.350pt}}
\put(850,158){\rule[-0.175pt]{0.350pt}{86.483pt}}
\put(850,158){\rule[-0.175pt]{141.167pt}{0.350pt}}
\sbox{\plotpoint}{\rule[-0.350pt]{0.700pt}{0.700pt}}%
\put(264,158){\usebox{\plotpoint}}
\put(264,158){\rule[-0.350pt]{282.335pt}{0.700pt}}
\end{picture}\end{center}
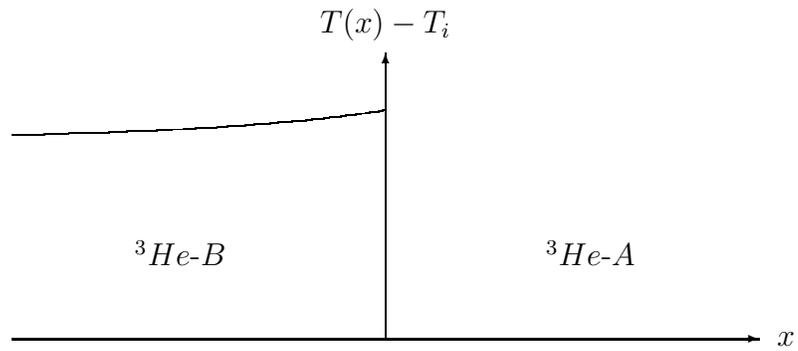
FIG.~2.\ \ The temperature field for $\dot u\gg c_2$, as in Eq.~(7a).

\begin{references}
\bibitem{boyd92} D. S. Buchanan, G. W. Swift, and J. C. Wheatley,
 Phys. Rev. Lett. {\bf 57}, 341 (1986);
 S. T. P. Boyd and G. W. Swift,
 Phys. Rev. Lett. {\bf 64}, 894 (1990);
 J. Low Temp. Phys. {\bf 86}, 325 (1992) and {\bf 87}, 35 (1992).
\bibitem{graliu90} M. Grabinski and M. Liu,
 Phys. Rev. Lett. {\bf 65}, 2666 (1990).
\bibitem{paliu92} P. Panzer and M. Liu,
 Phys. Rev. Lett. {\bf 69}, 3658 (1992);
 J. Low Temp. Phys. {\bf 92}, 127 (1993);
\bibitem{koliu93} P. Kost\"adt and M. Liu,
 Phys. Rev. Lett. {\bf 71}, 3513 (1993);
\bibitem{graliu88} B. Castaing and P. Nozi\'eres,
 J. Phys. (Paris) {\bf 41}, 701 (1980); H. Wiechert, J. Phys. C: Solid State
 Phys. {\bf 9}, 553 (1976); H. Wiechert and F. I. Buchholz, J. Low Temp. Phys.
 {\bf 39}, 623 (1980) and {\bf 51}, 291 (1993); S.E. Korshunov, Zh. Eksp.
 Teor. Fiz. {\bf 92}, 1320 (1987) [Sov. Phys. JETP {\bf 65}, 741 (1987)];
 M. Grabinski and M. Liu, J. Low Temp. Phys. {\bf 73}, 79 (1988);
\bibitem{seven} The non-equilibrium part of $Q$ vanishes only in the interface
 system, where we have $\6_xQ=0$ in the bulk, $\D Q=0$ at the interface, and
 $Q(x)=Q_{equil}$ for $x\gg|\la_2|$, hence $Q(x)=Q_{equil}$ also for
 $x\alt|\la_2|$. (For clarity, we confine the formula to the limit
 $\9u\ll c_2$.) In the laboratory system, we still have $\6_xQ=0$ in both
 phases, but the connecting condition is now $\D Q=-\r\9u\, l_{AB}$, and the
 above conclusion does not hold. Rather, the correct picture here is the
 heating up of the $A$ phase, within $\la_2$, by redistributing the latent
 heat $l_{AB}$ via second sound.
\bibitem{vw} D. Vollhardt and P. W\"olfe,
 {\it The Superfluid Phases of Helium 3}, Taylor and Francis, London (1990).
\bibitem{Legg} Obviously, the assumption $v_n=0$ in the original calculation
 of Leggett and Yip becomes valid for radii $R\ll 0.1$cm; cf.
 A. J. Leggett and S. K. Yip, in {\it Helium Three} (eds. W. P. Halperin and
 L. P. Pitaevskii), Modern Problems in Condensed Matter Sciences, Vol. 26,
 p. 523, North-Holland, Amsterdam (1990);
 A. J. Leggett, J. Low Temp. Phys. {\bf 87}, 571 (1992).
\end{references}
\end{document}